\documentclass[useAMS,usenatbib]{mn2e}
\usepackage{epsfig}

\def\ms{{\rm ms}}
\def\mb{{\rm mb}}
\def\in{{\rm in}}

\def\bh{{\rm H}}
\def\edd{{\rm Edd}}
\def\rcr{{\rm cr}}

\def\p{{\rm peak}}

\title[Accretion and Growth of Black Holes]
{Accretion, Growth of Supermassive Black Holes, and Feedback in Galaxy Mergers}
\author[Li-Xin Li]{Li-Xin Li\thanks{E-mail: lxl@pku.edu.cn}\\
Kavli Institute for Astronomy and Astrophysics (KIAA), Peking University, Beijing 100871, P. R. China}

\begin{document}

\date{}


\pagerange{\pageref{firstpage}--\pageref{lastpage}} \pubyear{2012}

\maketitle

\label{firstpage}

\begin{abstract}
Super-Eddington accretion is very efficient in growing the mass of a black 
hole: in a fraction of the Eddington time its mass can grow to an arbitrary 
large value if the feedback effect is not taken into account. However, since 
super-Eddington accretion has a very low radiation efficiency, people have 
argued against it as a major process for the growth of the black holes 
in quasars since observations have constrained the average accretion 
efficiency of the black holes in quasars to be $\ga 0.1$. In this paper
we show that the observational constraint does not need to be violated if
the black holes in quasars have undergone a two-phase growing process:
with a short super-Eddington accretion process they get their masses inflated 
by a very large factor until the feedback process becomes important, then 
with a prolonged sub-Eddington accretion process they have their masses 
increased by a factor $\ga 2$. The overall average efficiency of this two-phase 
process is then $\ga 0.1$, and the existence of black holes of $10^9
M_\odot$ by redshift $6$ is easily explained. Observational test of the 
existence of the super-Eddington accretion phase is briefly discussed.
\end{abstract}


\section{Introduction}
\label{intro}

The very existence of black holes of masses $10^9 M_\odot$ or more in quasars
at cosmic redshift $\ga 6$ \citep{fan01,barth03,wil03} has greatly challenged 
the theory for the growth of supermassive black holes. Recently, a $2\times
10^9M_\odot$ black hole has also been identified in a quasar at redshift $7$
\citep{mor11}. The most natural way for the growth of black holes is accretion 
of gases. However, as discussed below, it is very
unlikely that a black hole can grow to a mass of $10^9 M_\odot$ by redshift
$6$ by accretion from a standard geometrically thin accretion disk.

An astronomical black hole has two independent parameters, its mass $M_\bh$
and angular momentum $J_\bh$. A dimensionless spin parameter is defined by
$a_* \equiv c J_\bh/G M_\bh^2$, where $G$ is the Newtonian gravitational 
constant, and $c$ is the speed of light. According to general relativity,
$a_*$ must have a amplitude smaller than or equal to unity, i.e., $a_*^2 \le 
1$.

A geometrically thin disk is very efficient in converting mass into radiation
and spinning up a black hole. When the effect of photon recapture by the black 
hole is included, the maximum spin that a black hole can acquire by accretion 
from a thin accretion disk is $a_*=0.998$ (the ``canonical'' spin, Thorne 1974),
corresponding to a disk efficiency $\approx 0.3$ (the efficiency
in converting rest mass into radiated energy; see also Li 
et al. 2005). When a black hole is spun up from the non-rotating state ($a_*=
0$) to the canonical state ($a_* =0.998$), its mass grows by a factor of 
$\approx 2.7$. 

Thus, if a supermassive black hole had been growing by 
accretion from a thin disk, we would expect that its spin had already 
been equal to the canonical value from the very early stage, so that during 
the majority part of the history of the black hole growth the disk efficiency 
should be close to $0.3$. With such a high disk efficiency, for a black hole 
to grow to a mass of $10^9 M_\odot$ by the redshift $z=6$ through accretion 
from a thin disk, its initial mass should be $\ga 10^7 M_\odot$ (see 
Sec.~\ref{eqs}). This is almost impossible since among the seed black holes 
that have been proposed, the less exotic ones are in the range of $10^2$ -- 
$10^5 M_\odot$ \citep{vol05,vol10,ale11}.

For a black hole to grow from a reasonable seed mass to a mass of $10^9 
M_\odot$ by $z=6$ through accretion, the disk efficiency has to be very low. 
Although accretion modes with low efficiency exist (e.g., accretion through a 
geometrically thick disk), observations have provided stringent constraint 
on the average accretion efficiency. \citet{sol82} has proposed a simple 
approach to constrain the disk accretion efficiency in quasars: dividing 
the observed integrated luminosity density of quasars by the observed local 
black hole mass density. Based on this approach, \citet{yu02} found that the 
average disk accretion efficiency
for black holes in quasars must be $\ga 0.1$, otherwise the accreted mass 
density of the black holes during quasar phases would exceed the local mass 
density of black holes \citep[see also][]{elv02,wan06}. This constraint 
indicates that the geometrically 
thin disk accretion phase must have occupied a significant fraction of time
during the cosmic black hole accretion history.

Recently, disks misaligned with the black hole spin have been revisited by
\citet{kin05}. A major discovery is that when the disk angular momentum is less
than twice of the black hole spin and the angle between them 
satisfies certain condition the alignment torque between the black hole and 
the disk will counter-align the disk to the black hole spin. Then, it is 
possible that under some favorable conditions accretion is composed of a 
series of episodes with the disk angular momentum randomly orientated relative 
to the black hole spin so that spinning-up and spinning-down of the black hole 
have about equal possibilities. For example, \citet{kin08} have proposed that if
the disk is self-gravitating, it may repeatedly collapse producing accretion 
with a series of random orientations and thus the accretion will result in 
counter-alignment roughly half of time. Then, if the repeating process happens
very frequently, the spin parameter of the black hole will fluctuate around a 
mean value near zero, indicating a lower efficiency in converting accretion mass
into radiated energy than a maximally spinning black hole. For recent review 
and discussion on misaligned disks and their effects on evolution of the black
hole spin, see \citet{fan11}.

However, \citet{vol07b} argued that within the cosmological framework, 
where the most massive black holes have grown in mass via merger-driven 
accretion, 
one expects that disk accretion tends to make most supermassive black holes 
in elliptical galaxies to have on average higher spins than black holes in 
spiral galaxies. They proposed that the evolution of supermassive black holes
in elliptical galaxies are dominated by long and continuous accretion episodes
arisen from major mergers so their spins are brought up to very high values. 
While in spiral galaxies, growth of black holes are probably dominated by 
random and small accretion episodes (e.g., tidally disrupted stars, accretion 
of molecular clouds) and hence the black holes on average have low value spins.
They also argued that their models are in agreement with by the discovery 
that disk galaxies tend to be weaker radio sources than elliptical galaxies 
\citep{sik07}.
 
Even though it is possible that the black hole may not get a large net spin
through random accretion, it is still unlikely that accretion through 
geometrically thin disks is efficient enough to grow the mass of the black
hole to $>10^9 M_\odot$ by $z=6$. As will be shown in Sec. \ref{eqs}, for a
geometrically thin disk around a non-spinning black hole with a luminosity 
$0.3$ times the
Eddington luminosity, to have the black hole mass $3\times 10^9 M_\odot$ by 
$z=6$ the initial mass of the seed black hole must be at least $1.6\times
10^5 M_\odot$.

Growth of supermassive black holes by mergers has also been discussed in 
the literature \citep{kau00,hug03,sha05,ole06,ber08}. However, the black 
hole growth via merging is not expected to be very effective, especially when 
the recoil speed caused by the emission of gravitational wave is large 
\citep{ole06,vol07}.

Growth of black holes by super-Eddington accretion has been extensively 
explored \citep{kaw03,kaw04,beg06}. In particular, with Monte Carlo simulation 
\citet{vol05} have investigated a model for the early assembly of supermassive 
black holes by the merger tree of mini-halos, with a stable super-Eddington 
accretion phase at redshift $\ga 20$. They found that even a short phase of 
super-Eddington accretion eases the requirement by the existence of black 
holes of masses $\ga 10^9 M_\odot$ in quasars at redshift $\sim 6$. Recent study
on the growth of black holes by accretion and super-Eddington accretion includes
\citet{wan08}, \citet{wyi11}, and \citet{par12}.

Super-Eddington accretion, for which the mass accretion rate exceeds the 
Eddington limit but the luminosity does not, is very effective in feeding
a black hole but has a very low radiation efficiency. Hence, people have 
argued against super-Eddington accretion by the observational constraint 
on the average accretion efficiency $\ga 0.1$ \citep{yu02,elv02,wan06}. 

In this paper, we perform an analytical investigation of the growth of 
supermassive black holes via joint sub- and super-Eddington accretion in the 
framework of coevolution of black holes, quasars, and galaxies. This framework, 
in which quasar activity is assumed to be triggered by mergers of galaxies
\citep{mat05,hop05,hop06a,hop06,mal07,sij07,mat08,tre10,deb11}, has been 
motivated by the observations that black hole masses in nearby galaxies 
correlate with 
some properties of the host galaxies, such as the bulge luminosities and 
masses \citep{kor95,mag98}, and the central velocity dispersion 
\citep{fer00,geb00}.
However, we point out that the arguments in this paper in favor of 
super-Eddington accretion are valid for any mechanism that drives gases rapidly
toward the galactic nucleus and fuel the growth of the black hole. Galaxy 
mergers are only one of such mechanisms, other mechanisms include, for 
instance, global disk instabilities \citep{mo98,col00,bow06}.  

We assume that super-Eddington accretion takes place during a major merger 
of galaxies, since a major merger is expected to bring a lot of cold gas to 
the nucleus of the merged galaxy. We show that, with super-Eddington 
accretion the mass of a black hole grows very rapidly. Within a fraction of 
$10^8 {\rm yr}$, the mass of the black hole can grow to an arbitrary large 
value, provided that there is enough gas to accrete. We show that if the 
feedback effect is taken into account, a super-Eddington accretion phase will 
switch to a sub-Eddington phase when the mass of the black hole becomes large 
enough ($\ga 10^8 M_\odot$), then a black hole of mass $10^9 M_\odot$ can 
easily be produced by $z=6$. 

We argue that, since the super-Eddington phase can only take place in 
favorable conditions (e.g., during a major merger of galaxies), it can at most
occupy a short period of time during the history of a supermassive black 
hole. Then, the accretion efficiency averaged over the whole accretion 
history can easily reach a value $\ga 0.1$, consistent with the observational 
constraint.

The paper is organized as follows. In Sec.~\ref{eqs} we write down the 
fundamental equations governing the evolution of the mass and the spin of a 
black hole under disk accretion, and show that the standard thin disk accretion
is not efficient enough to grow a black hole. In Sec.~\ref{mass} we estimate 
the mass accretion rate with the Bondi model, and show that accretion will 
remain super-Eddington if it is initially so. In Sec.~\ref{sub}, we review the 
solution for sub-Eddington accretion. In Sec.~\ref{super}, we present the 
solution for super-Eddington accretion, and show that super-Eddington is very 
efficient for growing the black hole. In Sec.~\ref{twos} we propose a 
two-phase scenario 
for the growth of supermassive black holes: with an initial super-Eddington 
phase and a later sub-Eddington phase, the mass of the black hole grows rapidly
and the observational constraint on the average accretion efficiency is not 
violated. In Sec.~\ref{toy}, we present a simple model for the growth of
supermassive black holes in quasars, with the feedback effect being taken into
account. In Sec.~\ref{concl} we draw our conclusions and briefly discuss the
observational test of the existence of super-Eddington accretion phase in 
quasars and active galactic nuclei (AGNs).

We assume a flat $\Lambda$CDM cosmology with $\Omega_m=0.26$, $\Omega_\Lambda =
0.74$, and $H_0 =72$ km s$^{-1}$ Mpc$^{-1}$ \citep{dun09}.

\section{Growth of Black Holes via Accretion: Fundamental Equations}
\label{eqs}

Let us express the mass of the black hole, the mass accretion rate, and time in
dimensionless parameters
\begin{eqnarray}
	M_\bh = m M_{\bh,0} \;, \hspace{0.6cm}
	\dot M = \xi L_\edd /c^2 \;, \hspace{0.6cm}
	t = \tau t_\edd \;,  \nonumber
\end{eqnarray}
where $M_{\bh,0}$ is the initial mass of the black hole, $L_\edd = 1.3\times
10^{38} (M_\bh/M_\odot)$ erg s$^{-1}$ is the Eddington luminosity, and
\begin{eqnarray}
	t_\edd \equiv \frac{M_\bh c^2}{L_\edd} 
		= 4.51 \times 10^8\; {\rm year} \nonumber
\end{eqnarray}
is the Eddington time.

Note, our definition of the Eddington time, $t_\edd$, differs from the Salpeter
time by a factor $\varepsilon$, the efficiency of accretion. When $\varepsilon
=0.1$, $t_\edd$ is ten times of the Salpeter time. Although the Salpeter time
is more often used in the literature, we find that $t_\edd$ is more convenient,
at least for the purpose of the current paper.

Then, the evolution of the mass and the spin of the black hole under disk 
accretion is determined by \citep{tho74}
\begin{eqnarray}
	\frac{dm}{d\tau} &=& \xi m E_\in^\dagger \;, \label{dmdt} \\
	\frac{da_*}{d\tau} &=& \xi \left(\frac{L_\in^\dagger}{M_\bh}
		-2 a_* E_\in^\dagger\right) \;,
	\label{dadt}
\end{eqnarray}
where $E_\in^\dagger$ is the specific energy, $L_\in^\dagger$ is the specific
angular momentum of a disk particle at the inner boundary of the disk.

The efficiency of the disk in converting rest mass into energy is $\varepsilon 
= 1- E_\in^\dagger$. The luminosity of the disk is 
\begin{eqnarray}
        L = \varepsilon\dot{M} c^2 = \varepsilon\xi L_\edd \;.
	\label{lum}
\end{eqnarray}

In the theory of accretion disks it is usually assumed that the luminosity of
the disk is bounded by the Eddington luminosity. The arguments are that if the
luminosity of the disk radiation exceeds the Eddington luminosity, it is 
expected that the intense pressure of the disk radiation will halt the 
accretion flow so that steady accretion with luminosity greater than the 
Eddington luminosity is impossible \citep{fra02}. So, we assume that the 
luminosity is bounded by the Eddington luminosity, 
i.e.\footnote{Detailed calculations on supercritical accretion indicate that
when the mass accretion rate exceeds the critical mass accretion rate defined
by the Eddington luminosity (i.e., when $\xi>\xi_\rcr$, $\xi_\rcr$ is defined by
eq.~\ref{xi_cr}), the luminosity of the disk is not cut-off abruptly by 
the Eddington luminosity. Instead, above the critical mass accretion rate
the disk luminosity varies logarithmically with the mass accretion rate and 
about $10L_\edd$ can be approached for very high mass accretion rate 
\citep{abr88,wat06}.}
\begin{eqnarray}
        \varepsilon\xi\le 1 \;. \label{xi_con}
\end{eqnarray}

If the disk is geometrically thin, its particles move on Keplerian orbits in
the equatorial plane.\footnote{Even if initially the disk is not in
the equatorial plane, the Bardeen-Petterson effect will cause the inner region
of the disk to have its angular momentum to be aligned \citep{bar75} or
counter-aligned \citep{kin05} to the spin of the black hole quickly.} 
Then, the inner boundary of the disk is at the marginally stable circular 
orbit, i.e., $r_\in = r_\ms$ \citep{pag74}. The disk efficiency is then 
$\varepsilon_0
=\varepsilon(r_\ms) = 1-E_\ms^\dagger$. This efficiency is a function of the
spin of the black hole. When $a_* = 0$, the efficiency is $\varepsilon_0 
\approx 0.06$. When $a_* = 0.998$, the maximum spin that a black hole can get
through thin disk accretion, the efficiency is $\varepsilon_0 \approx 0.3$.

\begin{figure}
\vspace{2pt}
\includegraphics[angle=0,scale=0.69]{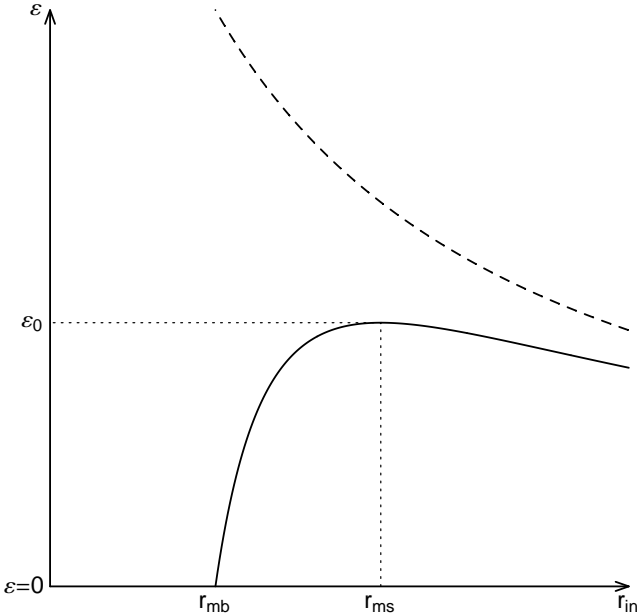}
\caption{The efficiency of a relativistic disk around a Kerr black hole
as a function of the radius of the disk inner boundary (solid line). 
The efficiency peaks at $r_\ms$, the marginally stable orbit, and drops 
to zero at $r_\mb$, the marginally bound orbit. For comparison, the 
efficiency of a Newtonian disk is shown with a dashed curve, which
monotonically increases as the radius of the disk inner boundary 
decreases.
}
\label{eff}
\end{figure}

Substituting $\varepsilon=\varepsilon_0$ into the inequality~(\ref{xi_con}),
we get a limit on the mass accretion rate for thin disk accretion: 
$\xi\la\xi_\rcr$, where
\begin{eqnarray}
        \xi_\rcr\equiv\frac{1}{\varepsilon_0} \;. \label{xi_cr}
\end{eqnarray}
Hence, a necessary condition for the disk to be geometrically thin is that 
$\xi<\xi_\rcr$ (i.e., the accretion rate must be sub-Eddington) since then 
the disk luminosity is below the Eddington 
limit.\footnote{Here we do not consider the exotic model of geometrically thick
disks with very low luminosity, like the ADAF model of \cite{nar94}. ADAF is
not expected to operate in the early stage of galaxy evolution (R. Narayan,
private communications).}

When $\xi>\xi_\rcr$, the disk has a super-Eddington accretion rate, and the 
inner region of the disk must be geometrically thick due to the trap of 
radiation \citep{beg79,abr80}. In this case, the inner boundary of the disk 
is located at a place between the marginally stable orbit and the marginally bound orbit 
(i.e., $r_\mb < r_\in<r_\ms$), and its efficiency is smaller than 
$\varepsilon_0$ so that the luminosity does not exceed the Eddington limit 
\citep{abr80,pac82a,pac82b}.

Even for a geometrically thick accretion disk, the specific energy and the
specific angular momentum of disk particles are still approximately equal to
the Keplerian values at the inner boundary of the disk since where the gas and
radiation pressure is always negligible, although this is not
the case for particles on orbits of larger radii. Thus, we always set 
$E_\in^\dagger$ and $L_\in^\dagger$ to their Keplerian values as though 
particles on the inner boundary move on a Keplerian circular orbit.

The specific energy $E^\dagger$ for particles on Keplerian circular orbit in the
equatorial plane of a Kerr black hole minimizes at the marginally stable orbit 
$r=r_\ms$, and $E^\dagger=1$ at the marginally bound orbit $r=r_\mb$ 
\citep{bar72}. Hence, $\varepsilon$ maximizes when the disk inner boundary is 
at the marginally stable orbit, and $\varepsilon =0$ at the marginally bound 
orbit (Fig.~\ref{eff}). For an efficiency between zero and the maximum value 
$\varepsilon_0$, the inner boundary of the disk is at a radius between $r_\mb$
and $r_\ms$. For comparison, in Fig.~\ref{eff} we also show the efficiency
of a Newtonian disk, which monotonically increases as the radius of the disk
inner boundary decreases.

For super-Eddington accretion ($\xi>\xi_\rcr$), the luminosity is $L=L_\edd$. 
Thus, the disk efficiency is $\varepsilon = 1/\xi$. In the case of extreme 
super-Eddington accretion with $\xi\gg 1$, the efficiency $\varepsilon\ll 1$ 
so we must have $r_\in \approx r_\mb$.

In study of the growth of supermassive black holes, people often assume that 
the disk is geometrically thin (so its inner boundary is at the marginally 
stable orbit), and the ratio of the disk luminosity to the Eddington
luminosity is a constant number smaller than but close to unity. That is, 
$\mu \equiv \varepsilon_0\xi = \mbox{constant}$, and $0.1\la\mu\la 1$. In
this case, equation~(\ref{dmdt}) becomes
\begin{eqnarray}
	\frac{d\ln m}{d\tau} = \frac{\mu(1-\varepsilon_0)}{\varepsilon_0}
		\;. \label{dmdt0} 
\end{eqnarray}

When $\varepsilon_0=\mbox{constant}$, the solution to equation~(\ref{dmdt0})
is
\begin{eqnarray}
	m(\tau) = \exp\left[\frac{\mu(1-\varepsilon_0)}{\varepsilon_0}\,
	        \tau\right] \;, \label{m_sol0}
\end{eqnarray}
where we have set $m=1$ at $\tau=0$. Under thin disk accretion the spin of the 
black hole will quickly grow to the canonical value $a_* = 0.998$ and then
is saturated at that vale, when the effect of photon recapture is considered
\citep{tho74,li05}. For a black hole to spin up from a non-rotating state
to the canonical state, the mass of the black hole need only increase by a
factor of about $2.7$. Hence, the most natural value for a constant 
$\varepsilon_0$
is the efficiency when the black hole is in the canonical state, which is 
$\varepsilon_0=\varepsilon_{\max} \approx 0.3$. Then, the solution becomes
$m(\tau) \approx e^{2.3\mu\tau}$.

At redshift $z=6$, the age of the universe is about $0.96$ Gyr, and $\tau\approx
2.1$. Thus, since $\mu\la1$, at $z=6$ we should have $m\la 130$. To explain the 
existence of a black hole of mass $3\times 10^9 M_\odot$ at $z=6$ \citep{wil03}
by thin disk accretion, the mass of the seed black hole would have to be 
$\ga 2\times 10^7 M_\odot$. This is almost impossible since among the seed black
holes that have been proposed the less exotic ones are in the range of
$10^2-10^5\odot$ \citep{vol05,vol10,ale11}.

For a geometrically thin disk to be a good approximation, the disk luminosity 
should not exceed 0.3 times the Eddington luminosity \citep{mcc06}. Take $\mu
=0.3$ and $\varepsilon_0=0.06$ (the efficiency for a non-spinning black hole), 
we get $m(\tau) \approx e^{4.7\tau}$. Then, to explain the existence of a black 
hole of mass $3\times 10^9 M_\odot$ at $z=6$ by the growth of black holes via
the model of a thin disk around an 
always-non-spinning black hole, the mass of the seed black hole would have 
to be $\sim 1.6\times 10^5 M_\odot$. This would also require a quite exotic 
high-mass seed black hole.

\section{The Mass Accretion Rate}
\label{mass}

The assumption $\mu = \varepsilon_0\xi = \mbox{constant}$ in the standard
treatment may not be correct in reality. In fact, the mass accretion rate is 
determined by the boundary condition of the accretion flow. A widely adopted
formula for estimation of the mass accretion rate is given by the Bondi 
rate \citep{bon44,bon52,nov73,sha83}
\begin{eqnarray}
	\dot{M}=4\pi\lambda_s \rho_\infty \frac{G^2 M_\bh^2}
		{c_{s,\infty}^3} \;,
	\nonumber
\end{eqnarray}
where $\rho_\infty$ and $c_{s,\infty}$ are, respectively, the mass density
and the sound speed at infinity, and $\lambda_s \sim 1$ is a number depending 
on the state equation of the gas.

The Bondi rate was derived from a steady and spherical accretion flow. For the 
case of disk accretion, the accretion flow cannot be spherically symmetric and
steady even at a distance far from the central black hole. \citet{hop10} have
shown that the disk accretion flow is highly nonsymmetric and the accretion 
rate is 
highly time variable for a given set of conditions in the galaxy at $\sim$ kpc.
Recently, \citet{hob12} have shown that when the gas is in a state of
free-fall at the evaluation radius due to efficient cooling and the dominant 
gravity of the surrounding halo, the Bondi model cannot give correct
estimation of the accretion rate. They have proposed an expression for the
sub-grid accretion rate which interpolates between the free-fall regime and the 
Bondi regime by taking into account the contribution of the halo to the gas
dynamics.

If $\rho_\infty$ and $c_{s,\infty}$ are constants in time, by the Bondi formula
we have $\dot{M} \propto M_\bh^2$. Since $L_\edd\propto M_\bh$, by the 
definition of $\xi$ we have $\xi \propto M_\bh \propto m$. The formula proposed
by \citet{hob12} takes into account the gravity of the halo, which results a
mass accretion rate that varies with $M_\bh$ slower than that given by the 
Bondi rate. Therefore, to make the results more general, we assume that
\begin{eqnarray}
	\xi = \xi_0 m^\gamma \;, \label{xi1}
\end{eqnarray}
where $0\le \gamma\le 1$, and $\xi_0$ is the value of $\xi$ at the beginning 
of accretion ($m=1$). The limit of $\gamma=1$ corresponds to the Bondi model. 
The limit of $\gamma=0$ corresponds to the case that the black hole accretes 
at a rate given by the Eddington rate multiplied by a constant number, which 
corresponds to a constant ratio of the disk luminosity to the Eddington 
luminosity when the disk efficiency $\varepsilon$ remains a constant. A 
value of $\gamma$ between 0 and 1 may represent a more realistic
case, for instance the model of \citet{hob12}.

Suppose the black hole is always spun up by accretion, i.e., $a_*>0$ does not 
decrease with time. Then, $\varepsilon_0$ does not decrease with time. If at 
the beginning the accretion is super-Eddington (i.e., $\xi_0>\xi_\rcr$), it 
will remain super-Eddington as accretion goes on as long as $\gamma\ge 0$, 
until some feedback effect becomes important to reduce the accretion rate, 
or the fuel is finally exhausted. Even if at the beginning the accretion is 
sub-Eddington (i.e., $\xi_0<\xi_\rcr$) it will eventually becomes 
super-Eddington if $\gamma>0$ provided that the accretion takes place for
a sufficiently long time.

It is under debate how much a fraction of the total inflowing gas is expelled
into an outflow when the total mass accretion rate is super-Eddington
\citep{sha73,ogi98,ogi01,bla99,lip99}. The concept of Eddington luminosity 
was derived by the consideration that when the luminosity exceeds the 
Eddington limit the intense pressure of the radiation on the
ionized gas would overcome the gravity of the central object onto which the
gas accreted so that the accretion would be halted until the luminosity
got below the Eddington limit. However, since $L=\varepsilon \dot{M} c^2$, the
reduce in the luminosity does not necessarily requires the reduce in the
mass accretion rate, it can also be realized by reducing the accretion 
efficiency $\varepsilon$. 

Even for spherical accretion, it was found that an intense outflow may not be
produced when the accretion rate exceeds the Eddington limit \citep{beg79}.
As the accreting gas falls deep enough in the gravitational well of the central
black hole so that the Eddington luminosity is approached, a ``trapping 
radius'' of radiation is crossed within which the radiation is trapped in 
the gas and dragged into the black hole by the inflowing gas. Effectively, 
the accretion efficiency is reduced while the mass accretion rate is not.
(For recent discussion on photon trapping, see Wyithe \& Loeb 2011.)

The situation is similar for the case of accretion disks \citep{abr80,abr88}. 
When the mass accretion rate exceeds the Eddington limit, a ``trapping radius'' 
is also formed in the inner region of the disk. Within the trapping radius the
disk radiation is trapped in the gas and dragged into the central black hole 
by the accreting gas, the effective accretion efficiency is hence reduced. In 
addition to this, general relativity also provides a new possibility for 
reducing the disk efficiency. Because of general relativity, the gravitational 
binding energy decreases as the inner boundary of the disk moves inward inside 
the marginally stable orbit, so that the disk efficiency is decreased 
(Fig.~\ref{eff}). As the inner boundary approaches the marginally bound orbit,
the disk efficiency approaches zero. Thus, to prevent the growth of the disk
luminosity, it is not necessary to reduce the mass accretion rate since it
can be realized by the photon trapping effect and/or the general relativistic 
effect. 

The ability of preventing the disk luminosity from exceeding the Eddington
luminosity by lowering the radiative efficiency without need of reducing the
mass rate accreting onto the central black hole leads to the possibility that 
in the black hole case a highly super-Eddington accretion may not necessarily 
lead to to a large outflow of mass. Therefore, in this paper we ignore the 
effect of outflows for super-Eddington accretion, in which case the inner 
boundary of the disk will move inward toward the marginally bound orbit to 
reduce the disk efficiency so that the luminosity does not exceeds the 
Eddington luminosity.

Substituting equation~(\ref{xi1}) into equations~(\ref{dmdt}) and (\ref{dadt}),
we obtain
\begin{eqnarray}
        \frac{dm}{d\tau} &=& \xi_0 m^{\gamma+1} E_\in^\dagger \;, 
	        \label{dmdt2} \\
        \frac{da_*}{d\tau} &=& \xi_0 m^\gamma \left(\frac{L_\in^\dagger}{M_\bh}
                -2 a_* E_\in^\dagger\right) \;.
        \label{dadt2}
\end{eqnarray}

\section{Sub-Eddington Accretion}
\label{sub}

Under sub-Eddington accretion, the disk is geometrically thin and the inner 
boundary is fixed at the marginally stable orbit: $r_\in=r_\ms$. Then, 
$E_\in^\dagger=E_\ms^\dagger $ and $L_\in^\dagger/M_\bh=L_\ms^\dagger/M_\bh$ are 
functions of $a_*$ only, equations~(\ref{dmdt2}) and (\ref{dadt2}) can be 
solved for $m(\tau)$ and $a_*(\tau)$.

The ratio of equation~(\ref{dmdt2}) and equation~(\ref{dadt2}) leads to a 
solution for $m(a_*)$
\begin{eqnarray}
	m(a_*) = \exp \left[\int_{a_*,0}^{a_*} \frac{E_\ms^\dagger da_*}
	          {L_\ms^\dagger/M_\bh -2a_* E_\ms^\dagger}\right] \;,
	\nonumber
\end{eqnarray}
which is independent of $\gamma$. With the solution of $m(a_*)$, we can solve 
for $\tau$ from (\ref{dadt2})
\begin{eqnarray}
	\tau(a_*) = \frac{1}{\xi_0}\int_{a_*,0}^{a_*} \frac{m(a_*)^{-\gamma}
	          da_*}{L_\ms^\dagger/M_\bh -2a_* E_\ms^\dagger} \;.
	\nonumber
\end{eqnarray}

The solution of $m(a_*)$ leads to that when a black hole is spun up from $a_*=
0$ to $a_*=0.998$, its mass is increased by a factor of $2.2$. This is somewhat
smaller than the value when the effect of photon recapture is considered, the
latter is about $2.7$ \citep{tho74}.

The solution of $\tau(a_*)$ shows that, for a black hole to spin up from $a_*=
0$ to $a_*=0.998$, it needs a time $\Delta t \approx 0.935\xi_0^{-1} t_\edd$ 
when $\gamma =0$, or $\Delta t\approx 0.635\xi_0^{-1} t_\edd$ when $\gamma =1$. 
If instead we assume that $\mu=\varepsilon_0\xi =\mbox{constant}$ during the 
accretion, we will have $\Delta t\approx 0.146\mu^{-1} t_\edd$. If the effect 
of photon recapture is considered, the value of $\Delta t$ should be somewhat
larger.

Therefore, by accretion the black hole is spun up to a limit spin $a_{*,\lim} 
= 0.998$ in a finite time, accreting a finite amount of mass. Afterwards, if 
the accretion rate remains sub-Eddington and the disk remains geometrically 
thin, $a_*$ will stay at $a_{*,\lim}$, and $E_\in^\dagger = E_{\lim}^{\dagger}
\equiv E_\ms^\dagger(a_{*,\lim}) \approx 0.7$. Then, the solution to 
equation~(\ref{dmdt2}) is
\begin{eqnarray}
	m(\tau) = \frac{1}{\left(1-\gamma\xi_0 E_{\lim}^{\dagger} 
	        \tau\right)^{1/\gamma}} \;,
	\label{m_sol1}
\end{eqnarray}
where we have set $m=1$ at $\tau=0$. Note, when $\gamma>0$, $m$ becomes
infinite at $\tau=1/\gamma\xi_0 E_{\lim}^\dagger$. 

However, the solution in equation~(\ref{m_sol1}) is valid only when $\xi < 
\xi_\rcr$, i.e., when the accretion is sub-Eddington. Since $\xi(\tau) = \xi_0 
m(\tau)^\gamma$, the accretion becomes super-Eddington (i.e., $\xi > \xi_\rcr$) 
at $\tau =\tau_s$, where
\begin{eqnarray}
	\tau_s = \frac{1-\xi_0\varepsilon_{\lim}}{\gamma\xi_0 
	       E_{\lim}^{\dagger}} \;,
	\nonumber
\end{eqnarray}
here $\varepsilon_{\lim} \equiv \varepsilon_0(a_{*,\lim}) \approx 0.3$. 

Hence, the solution in equation~(\ref{m_sol1}) holds only when $0<\tau<\tau_s$. 
By the time when the accretion becomes super-Eddington, the mass of the black 
hole has increased by a factor of $m=(\xi_\rcr/\xi_0)^{1/\gamma}$.

If at $\tau=0$ we have $\xi=\xi_0\ll 1$, then $\tau_s \approx 1/\gamma\xi_0 
E_{\lim}^{\dagger} \approx 1.4/\gamma\xi_0 \gg 1$, it would need a time much 
longer than the Eddington time to reach the super-Eddington phase. 
Hence, sub-Eddington accretion is very inefficient to grow the black hole.

As $\gamma\rightarrow 0$, the solution in equation~(\ref{m_sol1}) becomes that
in equation~(\ref{m_sol0}).

\section{Super-Eddington Accretion}
\label{super}

Under super-Eddington accretion, for which $\xi > \xi_\rcr$, the inner boundary 
of the disk is at a radius between $r_\mb$ and $r_\ms$. If the disk luminosity 
is limited by the Eddington luminosity, then $\varepsilon\xi=1$ (see the
discussion in Sec. \ref{eqs}), from which we have
\begin{eqnarray}
	E_\in^\dagger = 1- \frac{1}{\xi} \;.  \label{rin}
\end{eqnarray}
Equation~(\ref{rin}) determines the radius at the inner boundary of the disk,
for any given value of $\xi$.

Substituting equation~(\ref{rin}) into equation~(\ref{dmdt2}), we obtain
\begin{eqnarray}
	\frac{dm}{d\tau} = \xi_0 m^{\gamma+1} - m \;. \label{dmdt3}
\end{eqnarray}
The integration of equation~(\ref{dmdt3}) leads to
\begin{eqnarray}
	m(\tau) = \frac{1}{\left[\xi_0-(\xi_0-1)e^{\gamma\tau}
	        \right]^{1/\gamma}}
	\label{m_sol2}
\end{eqnarray}
for $0<\tau<\tau_\infty$, where
\begin{eqnarray}
	\tau_\infty \equiv \frac{1}{\gamma}\ln\frac{\xi_0}{\xi_0-1} \;.
	\label{t_infty2}
\end{eqnarray}

The mass $m$ becomes infinite at $\tau=\tau_\infty$. Thus, if $\gamma>0$, the 
black hole would have accreted an infinite amount of mass in a finite time.

Note, the solution in equation~(\ref{m_sol2}), as well as the value of 
$\tau_\infty$, does not explicitly depend on the initial spin of the black 
hole. The solution depends only on $\xi_0$ and $\gamma$.

For the accretion to be always super-Eddington, at $\tau=0$ we must have 
$\xi_0>1/\varepsilon_0$. Since $\varepsilon_0 <\varepsilon_0(a_*=1)
\approx 0.42$, we must have $\xi_0 > 2.4$. Thus, $\tau_\infty$ always 
satisfies the constraint
\begin{eqnarray}
	\tau_\infty < \frac{1}{\gamma}\ln \frac{1}{1-
	      \varepsilon_0(a_*=1)} \approx\frac{0.55}{\gamma}\;.
	\nonumber
\end{eqnarray}

In the limit $\xi\gg 1$, by equation~(\ref{rin}) we have $E_\in^\dagger\approx 
1$, so $r_\in \approx r_\mb$ and $L_\in^\dagger \approx 2 (r_\mb/M_\bh)^{1/2}$
\citep{bar72,abr80}. Equation~(\ref{dadt2}) then becomes
\begin{eqnarray}
	\frac{da_*}{d\tau} = 2\xi_0 m^\gamma \left[\left(\frac{r_\mb}{M_\bh}
		\right)^{1/2} - a_*\right] \;.
	\label{dadt3}
\end{eqnarray}
The equation for the mass is simplified to
\begin{eqnarray}
	\frac{dm}{d\tau} = \xi_0 m^{\gamma+1} \;.  \label{dmdt3a}
\end{eqnarray}

The solution to equation~(\ref{dmdt3a}) is
\begin{eqnarray}
	m(\tau) = \frac{1}{\left(1- \gamma\xi_0\tau\right)^{1/
	      \gamma}} \;, 
	\label{m_tau_large_xi}
\end{eqnarray}
which is consistent with the limit of equation~(\ref{m_sol2}) since then 
$\tau_\infty \approx 1/\gamma\xi_0 \ll 1$.

The ratio of equation~(\ref{dmdt3a}) and equation~(\ref{dadt3}) leads to
\begin{eqnarray}
	\int_1^m \frac{dm}{m} = \frac{1}{2} \int_{a_{*,0}}^{a_*} \frac{d a_*}
		{\left({r_\mb}/{M_\bh}\right)^{1/2} - a_*} \;,
\end{eqnarray}
whose solution is
\begin{eqnarray}
	m(a_*) = \exp\left[\frac{1}{2} \int_{a_{*,0}}^{a_*} \frac{d a_*}
		{\left({r_\mb}/{M_\bh}\right)^{1/2} - a_*}\right] \;,
        \label{m_a_super}
\end{eqnarray}
independent of $\gamma$.

With super-Eddington accretion, the effect of photon recapture on the evolution
of the black hole is not important since the radiation efficiency of the disk 
is very low. Then, the black hole can be spun up to a state with a spin 
arbitrarily close to unity \citep{abr80}. 

When $a_{*,0}=0$ and $a_* = 1$, 
by equation (\ref{m_a_super}) and the equation (2.19) of \citet{bar72} for
$r_\mb$, we have $m(1) = 2$. That is, when the black hole is spun up from 
$a_*=0$ to $a_*=1$ through super-Eddington accretion, its mass is doubled. In
the limit of $\xi\gg 1$ we have $\tau_\infty\approx1/\gamma\xi_0$, then by
equation (\ref{m_tau_large_xi}) the time needed for a black hole to spin up 
from $a_*=0$ to $a_*=1$ is $\approx (1-2^{-\gamma})\tau_\infty$.

We note that, although super-Eddington accretion in principle can
spin up the spin of the black hole all the way to a value arbitrarily close to
unity, the subsequent sub-Eddington accretion will lower the spin of the 
black hole to the canonical value $0.998$ by the effect of photon recapture.

\section{A Two-Phase Scenario for the Growth of Black Holes: the Feedback Effect}
\label{twos}

A main result in the last section is that the mass of a black hole can grow very
rapidly via super-Eddington accretion. Within a fraction of the Eddington time, 
the black hole would have taken all the mass of the surrounding gas. 

However, the growth of the black hole is limited by the feedback process 
\citep{sil98,hae98,fab99,kin03,kin05a}. When the mass of the black hole is 
sufficiently large, the dynamical effect of the disk radiation and/or outflow
on the surrounding gas will become important. The ambient gas will be swept 
away by the intense radiation and/or outflow from the central 
black hole, resulting that the mass accretion rate drops quickly. The accretion 
then transits from the super-Eddington phase to the sub-Eddington phase,
the growth of the black hole is slowed down, and finally stops when the 
fuel is exhausted.

The feedback process produced by the intense disk radiation becomes important 
when \citep{sil98}
\begin{eqnarray}
	M_\bh \ga 2\times 10^8 M_\odot \left(\frac{f}{0.05}\right)^{-1}
		\left(\frac{\sigma}{400\,{\rm km}\,{\rm s}^{-1}}\right)^5 \;,
	\nonumber
\end{eqnarray}
where $\sigma$ is the velocity dispersion of the halo hosting the galaxy, $f$ 
denotes the fraction of the black hole radiation converted to the kinetic 
energy of the gas. While in models where the outflowing gas can efficiently 
cool such that the flow is dominated by momentum, a different relationship
between $M_\bh$ and $\sigma$ was obtained \citep{kin03,kin05a}
\begin{eqnarray}
	M_\bh \sim 10^8 M_\odot \left(\frac{\sigma}{200\,{\rm km}\,{\rm s}^{-1}}
                \right)^4 \;.
	\nonumber
\end{eqnarray}
This relation is remarkably close to the observed $M_\bh-\sigma$ relation
\citep{tre02}. No matter which feedback model is more precise, the important 
point is that when the mass of the black hole becomes greater than a few 
$10^8 M_\odot$ the feedback process comes in to halt accretion.

Although super-Eddington accretion is very efficient in growing the black hole
mass, it has a very low efficiency in converting mass into radiation. The 
instant efficiency of super-Eddington accretion is simply $\varepsilon =1/\xi$, 
which is $\ll 1$ when $\xi\gg 1$. The average efficiency for a process of 
super-Eddington accretion is given by (see eq.~\ref{epsilon_av} below)
\begin{eqnarray}
	\overline{\varepsilon} = \frac{\int m d\tau}{\int \xi m d\tau} \;. 
	\nonumber
\end{eqnarray}
By equation (\ref{dmdt3}), in the limit $\xi\gg 1$ we get
\begin{eqnarray}
	\overline{\varepsilon} \approx \frac{1}{\xi} \times\left\{
	        \begin{array}{ll}
		        \ln m \;, &(\gamma=1) \;; \\
			\frac{1}{1-\gamma} \;, &(0<\gamma<1) \;;
		\end{array}
		\right.
	\nonumber
\end{eqnarray}
which is $\ll 1$.

However, this does not necessarily mean that the observational constraint 
$\overline{\varepsilon}\ga 0.1$ is violated. This is because of the following 
fact: super-Eddington accretion can only happen under very favorable 
conditions, in an environment with a lot of cold gas around the central
black hole, which could happen, e.g., during a major merger of galaxies. Hence, 
accretion in super-Eddington phase can at most occupy a short transient period 
(up to a fraction of the Eddington time) in the whole evolution history of a 
supermassive black hole. The rest of the history would be dominated by 
sub-Eddington accretion, which has a high efficiency in converting mass into
energy.

\begin{figure}
\vspace{2pt}
\includegraphics[angle=0,scale=0.515]{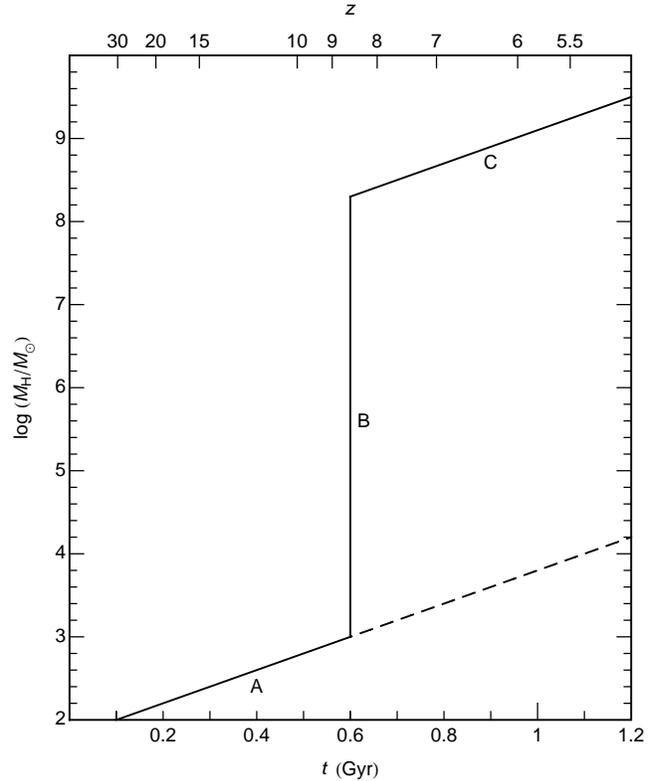}
\caption{A sketch for the possible accretion history of a supermassive
black hole (solid curve). At the cosmic time $t=0.1$ Gyr, a black hole of 
$100 M_\odot$ starts sub-Eddington accretion (A; $\mu=\varepsilon_0\xi = 
\mbox{constant}$). At $t=0.6$ Gyr, the black hole enters a phase of 
super-Eddington accretion (B; triggered by, e.g., a major merger of 
galaxies), within a short period of time ($\ll 1~{\rm Gyr}$) its mass
grows to $10^{8.3} M_\odot$. Then, the super-Eddington accretion ends (caused
by the feedback effect, presumably), and the accretion becomes sub-Eddington 
again (C). By $z=6$ ($t=0.96~{\rm Gyr}$) the black 
hole has a mass of $10^9 M_\odot$. The dashed curve denotes the extension 
of the initial sub-Eddington accretion, by which the mass of the black hole
 grows 
slowly. The horizontal axis is linear in time. The vertical axis is 
logarithm in the mass of the black hole. For references, the cosmic 
redshift from $z=5.5$ to $z=30$ is labeled on the top of the figure. 
}
\label{bh_grow}
\end{figure}

Therefore, we propose the following two-phase scenario for the growth of
a supermassive black hole. Imagine that at an early time of the universe,
e.g., at a redshift $z=20-30$, a seed black hole of $100 M_\odot$ was 
produced via collapse of a massive first generation star. The black hole
grew slowly by accreting mass from the surrounding gas. After a long
period of this primary sub-Eddington accretion stage, the mass of the
black increased to $10^3-10^4 M_\odot$ before the surrounding environment of
the black hole was changed suddenly caused by a major merger of galaxies.
The merger supplied a large amount of cold gas to the neighbor of the central 
black 
hole, and the black hole entered a phase of super-Eddington accretion. The 
mass of the black hole then grew very rapidly, until it was large enough 
that the feedback process took place to decrease the mass accretion rate and
slow down the growth of the black hole. The black hole then entered again
a phase of sub-Eddington accretion.

The super-Eddington phase took a very short period of time (e.g., $0.05 
t_\edd$ if at the beginning of the super-Eddington accretion we have $\xi = 
20$), but during which the mass of the black hole could inflate by a factor 
$\gg 1$. After the super-Eddington phase ended, the black hole 
continued growing slowly by sub-Eddington accretion. After a somewhat long 
time (e.g., several to ten Eddington time), the mass of the black hole 
increased by a factor $\zeta$, which is of order $1-10$. Although this 
second phase of sub-Eddington accretion is not very efficient for growing 
the mass of the black hole, it is crucial for boosting the average 
efficiency in converting mass into radiation. 

For the whole accretion process, the average efficiency in converting mass 
into energy is calculated by
\begin{eqnarray}
	\overline\varepsilon(t) = \frac{\int_{t_0}^t L dt}{c^2
		\int_{t_0}^t \dot{M} dt}
		= \frac{\int_{\tau_0}^\tau \varepsilon\xi m d\tau}
		{\int_{\tau_0}^\tau \xi m d\tau} \;,
	\label{epsilon_av}
\end{eqnarray}
where $L=\varepsilon\dot{M}c^2$ is the luminosity of the disk.

For the super-Eddington accretion phase which boosts the mass of the black hole
from $m=1$ to $m=m_1\gg 1$, by equation (\ref{dmdt3a}) it can be calculated 
that
\begin{eqnarray}
	\int\xi md\tau =\xi_0\int m^{\gamma+1}d\tau=\int dm = 
                m_1-1\approx m_1 \;, \nonumber
\end{eqnarray}
and by $\epsilon\xi=1$
\begin{eqnarray}
        \int\varepsilon\xi md\tau&=&\int md\tau =\frac{1}{\xi_0}\int m^{-\gamma}dm
                \nonumber\\
                &=&\frac{1}{(1-\gamma)\xi_0}\left(m_1^{1-\gamma}-1\right) 
                \approx\frac{m_1}{(1-\gamma)\xi_1}\;,
                \nonumber
\end{eqnarray}
where $\xi_1\equiv\xi_0m_1^\gamma\gg 1$.

For the sub-Eddington process which boosts the mass of the black hole from 
$m=m_1$ to $m=\zeta m_1$ with a constant efficiency $\varepsilon=\varepsilon_0
=1-E_\in^\dagger$, by equation (\ref{dmdt}) we have
\begin{eqnarray}
        \int\xi md\tau =\frac{1}{E_\in^\dagger}\int dm=\frac{(\zeta-1)m_1}
                {1-\varepsilon_0} \;, \nonumber
\end{eqnarray}
and
\begin{eqnarray}
        \int\varepsilon\xi md\tau=\varepsilon_0\int\xi md\tau=
                \frac{\varepsilon_0(\zeta-1)m_1}{1-\varepsilon_0} \;.\nonumber
\end{eqnarray}

Hence, for the entire accretion process we have
\begin{eqnarray}
        \int\xi md\tau \approx m_1+\frac{(\zeta-1)m_1} {1-\varepsilon_0}
                =\frac{\zeta-\varepsilon_0} {1-\varepsilon_0}m_1 \;,
                \nonumber
\end{eqnarray}
and
\begin{eqnarray}
        \int\varepsilon\xi md\tau\approx\frac{m_1}{(1-\gamma)\xi_1}+
                \frac{\varepsilon_0(\zeta-1)m_1}{1-\varepsilon_0}
                \approx\frac{\varepsilon_0(\zeta-1)}{1-\varepsilon_0}m_1 \;,
        \nonumber
\end{eqnarray}
since $\xi_1\gg 1$. Therefore the average efficiency during the entire history
of the black hole is
\begin{eqnarray}
	\overline{\varepsilon} \approx \frac{(\zeta-1)\varepsilon_0}
		{\zeta-\varepsilon_0} \;. \nonumber
\end{eqnarray}
Assume that during the sub-Eddington accretion phase the spin of the black hole
is kept at the canonical value and hence the radiation efficiency is 
$\varepsilon_0 \approx 0.3$. Then, if $\zeta>1.35$, we would have 
$\overline{\varepsilon} > 0.1$.

In this two-phase scenario for the cosmic growth of the black hole, the black 
hole gets weight through super-Eddington accretion, and does work through 
following-up sub-Eddington accretion. A sketch for a black hole to obtain 
a mass of $10^9 M_\odot$ this way by redshift $z=6$ is presented in 
Fig.~\ref{bh_grow}.

\section{A Simple Model for the Growth of Black Holes in Quasars}
\label{toy}

In this section we consider a simple model for the growth of black holes in
quasars. We assume that the mass accretion rate has the form $\dot{M}\propto 
M_\bh^2 e^{-M_\bh/M_\rcr}$, i.e., it is given by the Bondi rate when the mass of the
black hole is small and starts to decay exponentially when the mass of 
the black hole approaches a critical value $M_\rcr\sim 10^8M_\odot$. The decay 
in the mass accretion rate is assumed to be caused by the feedback effect
(see Sec. \ref{twos}). Then we have
\begin{eqnarray}
	\xi = \xi_0 m e^{-m/m_\rcr} \;, \label{xit}
\end{eqnarray}
where $m_\rcr\equiv M_\rcr/M_{\bh,0}$.

We adopt the slim disk model for super-Eddington accretion, where the disk
luminosity is $\propto \ln \dot{M} \propto \ln \dot{m}$ when $\xi \gg 1$
\citep{abr88,wat00,wat01}. To make a smooth transition from sub-Eddington to 
super-Eddington accretion, we assume a simple formula for the disk luminosity
\begin{eqnarray}
	L = \alpha^{-1} \varepsilon_0 L_\edd \ln \left(1+
		\alpha \xi\right) \;,
	\label{lum_slim}
\end{eqnarray}
where $\alpha$ is a constant, and $\varepsilon_0$ is the efficiency for a 
standard thin disk. When $\alpha\xi \ll 1$, equation~(\ref{lum_slim})
returns to the luminosity for a standard thin disk, $L \approx \varepsilon_0 
\xi L_\edd = \varepsilon_0\dot{M} c^2$. When $\alpha\xi\gg 1$, we have $L\approx
\alpha^{-1} \varepsilon_0 L_\edd \ln(\alpha \xi)$. Comparing this with the
results in \citet{wat00} and \citet{wat01}, we have $\alpha \sim 
\varepsilon_0/2 \approx 0.15$ if we assume that $\alpha_0\approx 0.3$ 
corresponding to the maximum efficiency of a standard thin disk.

By equation~(\ref{lum_slim}), the efficiency of the disk is
\begin{eqnarray}
	\varepsilon = \frac{\varepsilon_0}{\hat{\xi}} 
		\ln \left(1+\hat{\xi}\right) \;, \hspace{1cm}
		\hat{\xi} \equiv \alpha \xi \;.
\end{eqnarray}
Then, submitting $E_\in^\dagger = 1-\varepsilon$ into equation~(\ref{dmdt}), 
we get the equation that governs the evolution of the black hole mass
\begin{eqnarray}
	\frac{d m}{d\hat{\tau}} = m\left[\hat{\xi} - \varepsilon_0
		\ln \left(1+\hat{\xi}\right)\right] \;, \label{dmdt4}
\end{eqnarray}
where $\hat{\tau} \equiv \alpha^{-1}\tau$.

\begin{figure*}
\vspace{2pt}
\includegraphics[angle=0,scale=0.8]{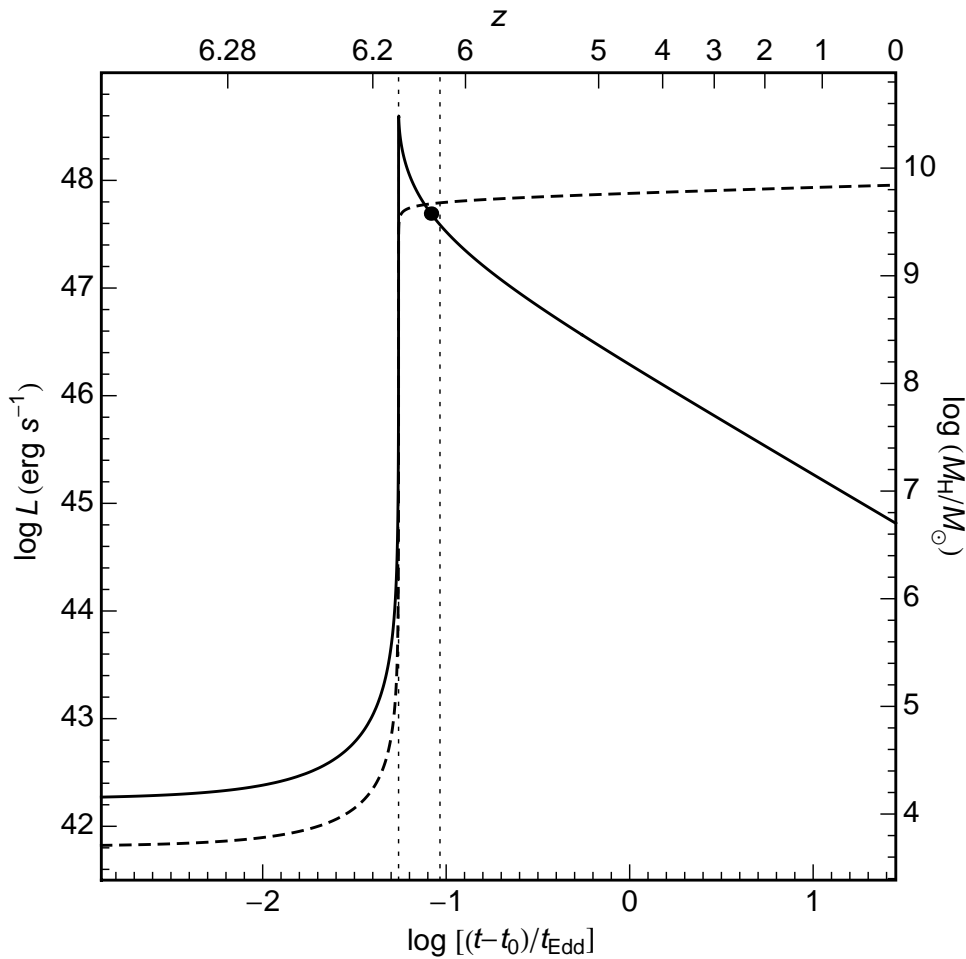}
\caption{Evolution of the luminosity (solid line) and the mass (dashed 
line) of the black hole, for the toy model in Sec.~\ref{toy}. The parameters
are: $\alpha=0.15$, $M_\bh(t_0) = 5,000 M_\odot$, $\xi(t_0)= 20$, and 
$M_\rcr = 3\times 10^8 M_\odot$. Super-Eddington accretion starts at 
$t_0=0.9$ Gyr, and ends at $t=0.937$ Gyr (marked by the black dot). Then 
the accretion becomes sub-Eddington, with an efficiency $\varepsilon_0=0.3$. 
The luminosity peaks at $t_\p=0.925$ Gyr. The two vertical
dotted lines bound the quasar epoch when the luminosity is above one-tenth 
of the peak luminosity. The time interval spanned by the two dashed lines 
is $1.7\times 10^7$ yr. On the top of the figure, the cosmic redshift is 
labeled for $z=0$, 1, 2, ... $6.28$ ($t_0=0.9$ Gyr corresponds to 
$z=6.296$).
}
\label{tm_fig1}
\end{figure*}

We set the parameters as follows: $\alpha=0.15$, $M_\rcr=3\times 10^8 M_\odot$. 
At the cosmic time $t_0=0.9$ Gyr, the initial mass of the black hole is 
$M_{\bh,0}=5,000 M_\odot$, and $\xi_0 = 20$. The black hole starts with a
super-Eddington phase with $\varepsilon_0\xi = 6$ and a luminosity $L \approx 
2.8 L_\edd$. Its mass evolves according to equation~(\ref{dmdt4}), 
supplemented by equation~(\ref{xit}) and $\hat{\xi} = \alpha\xi$. As the black 
hole grows the accretion becomes more super-Eddington since $\xi$ grows with 
time, until its mass becomes large enough ($>M_\rcr$), $\xi$ starts to 
decrease. When $\xi$ becomes smaller than $1/\varepsilon_0\approx 3.3$, the 
accretion enters the sub-Eddington phase during which the spin of the black 
hole is frozen at $a_*=0.998$, and the accretion efficiency is fixed at 
$\varepsilon_0=0.3$.

The numerical integration of equation~(\ref{dmdt4}) is shown in 
Fig.~\ref{tm_fig1}, where the dashed curve shows the evolution of the black
hole mass, the solid curve shows the evolution of the disk luminosity. The 
transition from super-Eddington to sub-Eddington takes place at $t = 0.937$
Gyr (the dark point in the figure), where $\varepsilon_0\xi =1$ and $L 
\approx 0.81 L_\edd$. During the super-Eddington phase (left to the dark 
point), the mass of the black hole grows by a factor of $\sim 10^6$. The 
luminosity of the disk peaks in the super-Eddington phase at $t_\p \approx
0.925$ Gyr where $\varepsilon_0\xi\approx 1.4\times 10^3$, with a peak
luminosity $L_\p \approx 13 L_\edd$. At the transition time the black hole 
has a mass of $4.7 \times 10^9 M_\odot$, and a luminosity of $4.9\times 
10^{47}$ erg s$^{-1}$. 

During the sub-Eddington phase (right to the dark point), the mass of the 
black hole grows very slowly, and the disk luminosity decays quickly due to 
the drop in the mass accretion rate. By the time of today ($z=0$), the mass
of the black hole would be $7.0\times 10^9 M_\odot$, and the luminosity of
the disk would be $6.5 \times 10^{44}$ erg s$^{-1}$.

The time-duration in which the luminosity of the disk exceeds one-tenth of 
the peak luminosity (the region bounded by the two vertical dotted lines in 
Fig.~\ref{tm_fig1}) is $\approx 1.7\times 10^7$ yr. This is in agreement with 
the observational constraint on the lifetime of unobscured quasars: $10^6 
{\rm yr} \la t_{\rm Q} \la 10^8 {\rm yr}$ \citep{mar04}.

\begin{figure}
\vspace{2pt}
\includegraphics[angle=0,scale=0.525]{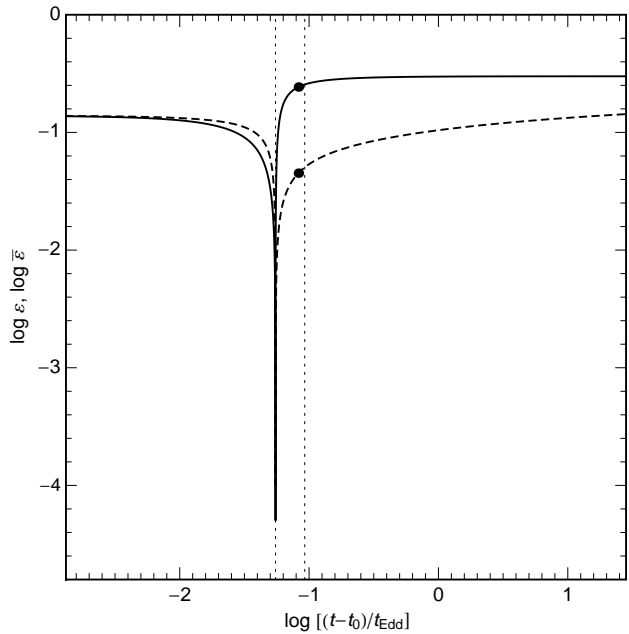}
\caption{Evolution of the instantaneous accretion efficiency (solid curve) and 
the average accretion efficiency (dashed curve) for the model in 
Fig.~\ref{tm_fig1}. The average efficiency is defined by the total energy 
emitted by the disk integrated from the start of the super-Eddington 
accretion (at $t_0=0.9$ Gyr) to a later time, divided by the total mass 
accreted during that time interval (eq.~\ref{epsilon_av}). As in 
Fig.~\ref{tm_fig1}, the black dots mark the transition from super-Eddington
to sub-Eddington accretion, and the two vertical dotted lines bound the 
quasar epoch.
}
\label{tm_fig2}
\end{figure}

In Fig.~\ref{tm_fig2} we show the instantaneous accretion efficiency (solid 
line) and the average accretion efficiency (dashed line), as a function of time.
The average accretion efficiency is defined by equation (\ref{epsilon_av}).

The instantaneous efficiency $\varepsilon$ minimizes at $t\approx0.92$ Gyr, 
when $M_\bh \approx M_\rcr$. When $M_\bh>M_\rcr$, the reduce in the mass 
accretion rate causes $\xi$ to drop quickly, and the efficiency $\varepsilon$ 
grows quickly. The average efficiency $\overline{\varepsilon}$ minimizes at 
$t\approx0.94$ Gyr, when $M_\bh \approx 8.1\times 10^8 M_\odot$. At the 
transition time, the instant efficiency becomes $0.24$, and the average 
efficiency is $0.045$. By the time of today, we have $\varepsilon\approx 0.3$
and $\overline{\varepsilon} \approx 0.14$. 

If in the definition of the average efficiency (eq.~\ref{epsilon_av}) the time
integration is over only the duration when the luminosity exceeds one-tenth of
the peak (i.e., over the region bounded by the two vertical dashed lines), 
then we obtain an average efficiency of $0.052$.

The overall average efficiency $\overline{\varepsilon} \approx 0.14$ is 
consistent with the observational constraints \citep{sol82,yu02,elv02,wan06}.

\section{Conclusions}
\label{concl}

We have proposed a simple model for the rapid growth of supermassive black 
holes in quasars. In this model, a major merger of two galaxies (or some
secular processes like global disk instabilities) brings a 
large quantity of cold gas to the region around the central black hole formed by
the coalescence of the two black holes in the nuclei of parent galaxies. 
Due to the
very high density of the cold gas, a super-Eddington accretion phase is turned
on. During this super-Eddington phase, the mass of the black hole inflates
quickly, without the feedback effect all the mass of the gas would have
been swallowed in a fraction of the Eddington time. The existence of black 
holes of masses $\sim 10^9 M_\odot$ at $z\sim 6$ can easily be explained 
with this model.

When the mass of the black hole becomes large enough ($\ga 10^8 M_\odot$), the
feedback effect becomes important which shuts off the super-Eddington 
accretion. Afterwards, the black hole enters a sub-Eddington accretion phase,
during which the mass of the black hole continues increasing but with a slower
rate, until all of the surrounding gas are taken over or a next event of 
merger/instability happens.

Although the super-Eddington phase has a very low efficiency in converting 
mass into radiation, the subsequent sub-Eddington phase has a very high
efficiency ($\approx 0.3$ for a standard thin disk around a nearly
maximal-spinning black hole). If during the later sub-Eddington phase the
mass of the black hole gets boosted by a factor $\ga 1.5$ then the overall 
average efficiency will be $\ga 0.1$, consistent with the
observational constraint on the average accretion efficiency of quasars
\citep{yu02,elv02,wan06}.

In our treatment of super-Eddington accretion we have ignored the effect of 
disk outflows. The judgment for our choice is based on the
fact that to prevent the disk luminosity from exceeding the Eddington 
luminosity the mass accretion rate does not have to be reduced since the
disk can adjust itself to have a low efficiency either through photon
trapping in the inner region of the disk, and/or through pushing the inner 
boundary of the disk toward the marginally bound orbit. Of course this is
an issue under debate and further detailed investigation is needed to 
determine if a strong outflow will be driven by a super-Eddington accretion
so that the majority of the gas accreting onto the black hole will be 
expelled to infinity by the radiation pressure. However, a recent study by 
\citet{wyi11} indicates that this will not happen at least in certain 
circumstances, where the photon
diffusion is too slow to expel the accreting gas.  

The scenario that we have proposed is consistent with the popular view that 
galaxy mergers, quasars, and the growth of supermassive black holes 
coevolve \citep{mat05,hop05,hop06a,hop06,sij07,mat08,tre10,deb11}.

It would be interesting to test the scenario with observations. Observational
evidence for super-Eddington accretion in quasars has been discussed by
\citet{col02}. Recently, \citet{kaw11} proposed a new method to explore 
super-Eddington accretion in AGNs by near-infrared 
observations. They found that generally the ratio of the AGN IR luminosity and 
the disk bolometric luminosity for super-Eddington AGNs is much smaller than 
that for sub-Eddington AGNs, caused by the self-occultation effect of the 
super-Eddington accretion flow. While for nearby galaxies currently undergoing 
major mergers, super-Eddington may be observed by direct or indirect 
measurements of the mass accretion rate and the luminosity like in the case
of Narrow-Line Seyfert 1 galaxies \citep{kaw03}.

\section*{Acknowledgments}

The author acknowledges R. Sunyaev for helpful comments and discussion when 
part of the work was being finished in Max-Planck-Institut f\"ur Astrophysik. 
He also thanks the referee R. Hickox for a very thorough and constructive 
report. This work was supported by the NSFC grants programme (no. 10973003) 
and the National Basic 
Research Programme of China (973 Programme under grant no. 2009CB24901).

\bsp

\label{lastpage}

\end{document}